\documentclass[11pt]{article}

\usepackage{color}
\usepackage{latexsym}
\usepackage{amssymb}
\usepackage{graphicx}

\newtheorem{Theorem}{Theorem}[part]

\newtheorem{Proposition}{Proposition}[part]

\newtheorem{Lemma}{Lemma}[part]

\newtheorem{Remark}{Remark}[part]

\renewcommand{\theDefinition}{\thesection.\arabic{Definition}}
\renewcommand{\theTheorem}{\thesection.\arabic{Theorem}}

\renewcommand{\theLemma}{\thesection.\arabic{Lemma}}
\renewcommand{\theRemark}{\thesection.\arabic{Remark}}

\renewcommand{\theequation}{\thesection.\arabic{equation}}

\def \R{\mathbb{R}}

\def \E{\mathbb{E}}
\def \F{\mathbb{F}}
\def \G{\mathbb{G}}

\def \P{\mathbb{P}}
\def \Q{\mathbb{Q}}

\def \Ac{{\cal A}}

\def \Cc{{\cal C}}
\def \Dc{{\cal D}}
\def \Ec{{\cal E}}
\def \Fc{{\cal F}}
\def \Gc{{\cal G}}

\def \Lc{{\cal L}}

 \def \Nc{{\cal N}}

\def \Tc{{\cal T}}

\def \Xc{{\cal X}}

\def \ep{\hbox{ }\hfill$\Box$}

\def\reff#1{{\rm(\ref{#1})}}

\def\beqs{\begin{eqnarray*}}
\def\enqs{\end{eqnarray*}}
\def\beq{\begin{eqnarray}}
\def\enq{\end{eqnarray}}

\begin{document}
\title{Optimal investment  on finite horizon \\  with random discrete order flow  in illiquid markets
\thanks{H. Pham would like to thank Peter Leoni (Electrabel) for interesting discussions on this illiquidity market modelling.}}
\author{Paul Gassiat\footnote{LPMA, University Paris 7, pgassiat@math.jussieu.fr }~~~
Huy\^en Pham\footnote{LPMA, University Paris 7, CREST, and Institut Universitaire de France, pham@math.jussieu.fr}~~~   
 Mihai S\^{\i}rbu\footnote{Department of Mathematics, University of Texas, Austin, sirbu@math.utexas.edu}
}

\maketitle


\begin{abstract}
We study the problem of optimal portfolio selection in an illiquid market with discrete order flow. In this market, bids and offers are not available 
at any time but trading occurs more frequently near a terminal horizon. The investor  can observe and trade the risky asset only at  exogenous random times corresponding to the order flow given by an inhomogenous Poisson process. By using a direct dynamic programming  approach, we first derive and solve 
 the  fixed point dynamic programming equation 
 and then perform a verification argument which provides the existence and characterization 
of optimal trading strategies. We prove the convergence of the optimal performance, when the deterministic 
intensity of the order flow approaches infinity at any time, 
to the optimal expected utility for an investor trading continuously in a perfectly liquid market model with no-short sale constraints. 
\end{abstract}

\vspace{7mm}

\noindent {\bf Key words:} liquidity modelling, discrete order flow, optimal investment, inhomogenous Poisson process, dynamic programming. 

 \vspace{7mm}

\noindent {\bf MSC Classification (2000)}:  60G51, 49K22, 93E20, 91B28.  

 \vspace{7mm}

\noindent {\bf  JEL Classification}:  G11,  C61. 

\newpage

\section{Introduction}

\setcounter{equation}{0} \setcounter{Assumption}{0}
\setcounter{Theorem}{0} \setcounter{Proposition}{0}
\setcounter{Corollary}{0} \setcounter{Lemma}{0}
\setcounter{Definition}{0} \setcounter{Remark}{0}

 Financial modeling very often relies on the assumption of continuous-time trading.
 This is essentially due to  the availability of the powerful tool of stochastic integration, which allows for an elegant representation of continuous-time trading strategies, and the analytic  tractability of the stochastic calculus,  typically illustrated by It\^o's formula.   
Sometimes this  assumption may not be very realistic in practice: illiquid markets provide a prime example. 
Indeed, an important aspect  of market liquidity is  the time  restriction both on trading and observation of the assets. For example, in power markets,  
trading occurs through a brokered OTC market, and the liquidity is really thin.  There could be  a possible lack of counterparty for a given order: bids and offers are not available at any time, and may arrive  randomly, while the investor can observe the asset only at these arrival times.   
Moreover, in these markets, because of the physical nature of the underlying asset, trading activity is really low far from the delivery, and 
is higher near the delivery.

In this paper, we propose a framework that takes into account such liquidity features by considering a discrete order flow. In our model, the investor can observe and trade over a finite horizon only at random times given  by an inhomogenous  
Poisson process encoding the quotes in this illiquid market.  To capture the high frequency of trading in the neighborhood of  the finite horizon,  we assume that the deterministic intensity of this inhomogenous Poisson process approaches infinity  as time gets closer to  the finite horizon.  In this context,  the objective of   the agent is to  maximize his/her expected utility from terminal wealth.

 Optimal investment problems with random discrete trading dates were studied by se\-veral authors.  Rogers and Zane \cite{rogzan02} and Matsumoto \cite{mat06} considered  trading times associated to the jump times of a Poisson process, but assumed  that price process is observed continuously, so that trading strategies are actually in continuous-time.  Recently,  Pham and Tankov  \cite{phatan08}  
 (see also \cite{cregozphatan08})  investigated an optimal portfolio/consumption choice problem  over an infinite horizon, where the asset  price,
 essentially extracted  from a Black-Scholes model, can be observed and traded only at the  random times  corresponding to a Poisson process with constant intensity.

  As mentioned above, here we investigate an optimal investment problem over a finite horizon with random trading and observation times  given by an inhomogeneous Poisson process with a deterministic intensity  going to infinity near the final horizon. 
 The underlying continuous-time asset  price  is  given by  an  inhomogeneous  L\'evy process.  We use a direct dynamic programming approach for solving this  portfolio selection problem. We first derive the fixed point dynamic programming equation (DPE) and provide a constructive proof for the existence of a solution to this  DPE in a suitable functional space by means of an iterative procedure. 
 Then, by proving a verification theorem, we obtain the existence and characterization of optimal policies.   
We also provide an approximation of the optimal strategies that involves only a finite number of iterations.
Finally, we address the natural question of convergence of our optimal investment strategy/expected utility when the arrival intensity rate becomes large at all times. 
We prove that the value function converges to the value function  of an agent who can trade continuously in a perfectly liquid market with no-short sale constraints.  A related  convergence result  was recently obtained by Kardaras and Platen \cite{KardarasPlaten} by considering continuous-time trading strategies  approximated by simple trading strategies with constraints, but with asset prices observed continuously. 
 Here we face  some additional subtleties induced by the discrete observation filtrations: the illiquid market investor has less information coming from observing the asset, compared to the continuous-time investor, but he/she has the additional information coming from the arrival times,  which is lacking in the perfectly liquid case.

The rest of the paper is organized as follows.  Section 2   describes the illiquid market model with the restriction on the trading times, and 
sets up  all the assumptions of the model. We formulate in Section 3 the optimal investment problem, and solve it by a dynamic programming approach and verification argument. Finally,  Section 4 is devoted to the convergence issue when the deterministic 
intensity of arrivals is very large at all times.

\section{The illiquid market model and  trading strategies }

\setcounter{equation}{0} \setcounter{Assumption}{0}
\setcounter{Theorem}{0} \setcounter{Proposition}{0}
\setcounter{Corollary}{0} \setcounter{Lemma}{0}
\setcounter{Definition}{0} \setcounter{Remark}{0}

We consider an illiquid market in which an investor can trade a risky asset over a finite horizon. In this market, bids and offers 
are not available at any time,  but trading occurs more frequently near the horizon. This is typically the case in power markets with forward contracts.  This market illiquidity feature is modelled  by   assuming that  the arrivals of buy/sell orders  occur at the jumps of an inhomogeneous Poisson process  with an increasing  deterministic intensity  converging to infinity  at the final horizon.  In order to obtain an analytically  tractable model, we further assume that the discrete-time observed asset prices come from an unobserved continuous-time stochastic process, which is independent of the sequence of  arrival times.  We may think about the continuous time process as an asset price process  based on fundamentals independent of time-illiquidity, which would be actually observed if trading occurred at all times.

More precisely, fixing  a probability space $(\Omega,\Gc,\P)$ and a finite horizon $T$ $<$ $\infty$,
 we consider  the fundamental  unobserved positive asset price $(S_t)_{0\leq t\leq T}$.  
An investor can observe and trade the asset \emph{only} at  some exogenous random times $(\tau_n)_{n\geq 0}$, $\tau_0$ $=$ $0$, such that $(\tau _n)_{n\geq 0}$ and $(S_t)_{0\leq t\leq T}$ are independent under the physical probability measure $\P$.

In order to obtain a stochastic control problem of Markov type for the utility maximization problem below, we assume an exponential-L\'evy  structure and some regularity/integrability conditions on  the continuous time positive price process $S$.    More precisely, we assume that 
\beqs
S_t =\mathcal{E}(L)_t,\ \ \ \ \ 0\leq t\leq T,
\enqs
 where the process $(L_t)_{0\leq t\leq T}$  is a semimartingale on $(\Omega, \Gc, \P)$ with 
 independent increments and jumps strictly greater than one. We use  $\mathcal{E}(L)$ to denote  the Dol\'eans-Dade stochastic exponential of $L$. The assumption  $\Delta L>-1$ ensures that   the asset $S$, as well as its left-limit $S_{-}$, are strictly positive at all times. 
It is well known that a semimartingale with independent increments has \emph{deterministic predictable characteristics}, see e.g. \cite{jacshi03}.  
Denoting by $\mu (dt,dx,\omega)$ the jump measure and by $\nu (dt, dx)$ its deterministic compensator, we assume that the large jumps are integrable, namely that 
the compensator measure $\nu(dt,dy)$ of $\mu$ satisfies
\beq \label{int-large-jumps}\int _0^T\int _{-1}^{\infty}y\,\nu (dt,dy)<\infty.
\enq
The L\'evy-Khintchin-It\^o decomposition implies that
\beqs L_t=A(t)+M_t+ \int _0^t\int _{-1}^{\infty}y (\mu (dt,dy)-\nu (dt,dy)),\ \ 0\leq t\leq T, 
\enqs
where  $A$ is a deterministic function of bounded variation and $M$ is a continuous local martingale with deterministic quadratic variation, so it is Gaussian. We further assume the following regularity of the deterministic predictable characteristics
\begin{enumerate}

\item there exists a  function $b:[0,T]\rightarrow \mathbb{R}$ such that 
\beq \label{reg-drift}\int _0^T|b(u)|du <\infty \textrm{~and~} A(t)= \int _0^tb(u)\,du,\ \ \ 0\leq t\leq T
\enq

\item there exists a function  $c:[0,T]\rightarrow (0, \infty) $ such that
\beq \label{reg-vol} \int _0^Tc^2(u)du<\infty \textrm{~and~} \langle M\rangle _t=\int _0^t c^2(u)\, du, \ \ \ 0\leq t \leq T.
\enq

\end{enumerate}
Conditions \reff{int-large-jumps}, \reff{reg-drift} and \reff{reg-vol} together mean that we actually make the following assumption on $L$:

\noindent {\bf (HL)}  \beqs L_t=\int _0^t b(u)\,du+\int _0^t c(u)\,dB_u+\int _0^t\int _{-1}^{\infty}y (\mu (dt,dy)-\nu (dt,dy)),\ \ 0\leq t\leq T,
\enqs
where $(B_t)_{0\leq T}$ is  a Brownian motion on $(\Omega, \mathcal{G},\mathbb{P})$  independent on the jump measure $\mu$,  the measure $\mu$ satisies 
\reff{int-large-jumps} and $b,c$ satisfy \reff{reg-drift} and \reff{reg-vol}.
We denote by 
\beqs
Z_{t,s} &=&\frac{S_s - S_t}{S_t}=\left \{e^{(L_s-L_t-\frac{1}{2}\int _t^s c^2(u)\,du)}\prod_{t<u\leq s}e^{-\Delta L_u}(1+\Delta L_u) \right \} -1, \;\;\;\;\; 0\leq t\leq s\leq T, 
\enqs
the return between times $t$ and $s$ (if trading is allowed at both times) and denote by 
\beqs
p(t,s,dz) &=& \P[Z_{t,s} \in dz] 
\enqs
the distribution of the return.
\begin{Remark}\label{HZ} {\rm Assumption ${\bf (HL)}$  ensures that

\noindent (i)  For all $0\leq t<s\leq T$ the  distribution $p(t,s,dz)$ has support whose interior is equal to $(-1,\infty)$.  

\noindent (ii) There exists some positive constant $C$ $>$ $0$ such that  $ \int_{(-1,\infty)} |z| p(t,s,dz)$ $\leq$ $C$, for all $0\leq t\leq s<T$. This means that, the expectation of the absolute value of the return is uniformly bounded in $(t,s)$, i.e. $\mathbb{E}[|Z_{t,s}|]\leq C$. The constant $C$ can be chosen as
$C=e^{\int _0^T|b(u)|du}$, where $b$ is defined in \reff{reg-drift}.

\noindent (iii) Since in our model the asset $S$  is not observed at terminal time $T$, there is no loss of generality if we assume that $S_T=S_{T-}$, which can be translated in terms of predictable characteristics as
$\nu (\{T\}, (-1, \infty))=0$. We will make this assumption for the rest of the paper.} 
\end{Remark}

 The sequence of observation/trading times is represented by  the jumps of an inhomogeneous (and independent on $S$) Poisson process $(N_t)_{t\in [0,T]}$  
with deterministic  intensity function $t$ $\in$ $[0,T)$ $\rightarrow$ 
$\lambda(t)$ $\in$ $(0,\infty)$, such that:
\beq \label{hypintens}
\int _0^t\lambda (u)du \; < \;  \infty,\ (\forall)\  0\leq t<T   & \mbox{ and } & \int_0^T \lambda(u) du \; = \;  \infty.  
\enq
 The simplest way to actually define such an inhomogeneous Poisson process is to consider  a homogeneous Poisson process $M$ with intensity equal to one, independent of $S$ and define
\beq 
\label{representation-N}
N_t=M_{\int _0^t \lambda (u)du} &\mbox{for} &\ \ 0\leq t<T.
\enq
Condition (\ref{hypintens})  ensures that the probability of having no jumps between any interval $[t,T]$, $t$ $<$ $T$, is null, and so  
the sequence $(\tau_n)$ converges increasingly to $T$ almost surely when  $n$ goes to infinity.  We also know  that the process of jump times 
$(\tau_n)_{n\geq 0}$ is a homogeneous Markov chain on $[0,T)$, and its  transition probability  admits a density given by: 
\beq \label{loitaun}
\P[ \tau_{n+1} \in ds | \tau_{n} = t] &=& \lambda(s) e^{-\int_t^s \lambda(u) du} 1_{\{t\leq s< T\}} \; ds, 
\enq
(which does not depend on  $n$).

An investor trading in this market can only observe/trade the asset $S$ at the discrete arrival times $\tau _n$. Therefore, the only information he/she has is coming from observing the two-dimensional process $(\tau _n, S_{\tau_n})_{n\geq 0}$. 
Taking this into account,  we introduce the discrete observation filtration  $\F$ $=$ $(\Fc_n)_{n\geq 0}$, with $\Fc_0$ trivial and  
\beq \label{F}
\Fc_n &=& \sigma\Big\{ (\tau_k,Z_k): 1\leq k \leq n\Big\}, \;\;\; n \geq 1, 
\enq
where we  denote by 
\beqs
Z_n=Z_{{\tau_{n-1}},\tau _{n}}, \;\;\;\;\; n\geq 1,
\enqs
the observed return process valued in $(-1,\infty)$.
\begin{Remark}\label{rem:conditional-distribution}
{\rm Assumption ${\bf (HL)}$  together with the independence of $S$ and $N$ ensures that  for all $n$ $\geq$ $0$, 
the (regular) distribution of $(\tau _{n+1}, Z_{n+1})$ conditioned on $\mathcal{F}_n$ is given as follows:
\begin{enumerate}
\item $
\P[ \tau_{n+1} \in ds | \mathcal{F}_n] =\lambda(s) e^{-\int_{\tau _n}^s \lambda(u) du}ds$
\item further conditioning on knowing  the next arrival time $\tau _{n+1}$, the return $Z_{n+1}$ has distribution
$$\P[Z_{n+1}\in dz|\mathcal{F}_n \vee \sigma (\tau _{n+1})]=p(\tau _n, \tau _{n+1}, dz).$$
\end{enumerate}
 }
 \end{Remark}

In this model a (simple) trading strategy is a real-valued  $\F$-adapted  process $\alpha$ $=$ $(\alpha_n)_{n\geq 0}$,   where 
$\alpha_n$ represents the amount  invested  in the stock  over the period $(\tau_{n},\tau_{n+1}]$ after observing the stock  price at time 
$\tau_{n}$.  Assuming that the money market pays zero interest rate, the  observed wealth process $(X_{\tau_n})_{n\geq 0}$ associated to a trading strategy $\alpha$ is governed by:
\beq \label{selfX}
X_{\tau_{n+1}} &=& X_{\tau_{n}} + \alpha_{n} Z_{n+1}, \;\;\; n \geq 0,
\enq
where $X_0$ is the initial capital of the investor. In order to simplify notation, we fix once and for all an  initial capital $X_0$ $>$ $0$ and   denote by $\Ac$ the set of trading strategies 
$\alpha$ such that  the wealth process stays nonnegative:
\beq \label{xtaunpos}
X_{\tau _n} &\geq& 0, \;\;\; n \geq 1. 
\enq
For the rest of the paper, we will call \emph{simple} these trading strategies where trading occurs only at the discrete times $\tau _n$, $n\geq 0$.

\begin{Remark} \label{remmarX}
{\rm {\it Constrained strategies.} From \reff{selfX} and the support property of $Z_{t,s}$ in Remark \ref{HZ} (i), the admissibility condition 
\reff{xtaunpos} on  $\alpha$ $\in$ $\Ac$ means that we have a  no-short sale  constraint (on both the risky and savings account asset): 
\beq \label{condadmiss}
 0 \; \leq \; \alpha_{n}  \;  \leq  \; X_{\tau_n},   \;\;\; \mbox{ for all } \;  n \geq 0. 
\enq
Moreover, since $Z_n$ $>$ $-1$ a.s. for all $n$ $\geq$ $1$,  the wealth process associated to $\alpha$ $\in$ $\Ac_0$ is actually strictly positive: 
\beqs
X_{\tau _n} &>& 0, \;\;\; n \geq 0. 
\enqs
}
\end{Remark}
For technical reasons, some related to the asymptotic behavior in Section \ref{asymptotic}, we need to define some continuous time filtrations along with the discrete filtration $\F$. To avoid confusion, we will denote   by $\mathbb{G}$ (with different parameters) all  continuous-time filtrations.  In this spirit, we define  the  filtration   $\G$ $=$ $(\Gc_t)_{0\leq t\leq T}$ generated by observing continously the process $S$ and the arrival times as
\beq \label{cont-filtration}
\Gc_t &=& \sigma  \big\{ (S_u, N_u), 0\leq u\leq t\} \vee \mathcal{N}, \ \ \ 0\leq t\leq T,
\enq
where  $N$ is the inhomogenous Poisson process in \reff{representation-N} and $\mathcal{N}$ are all the null sets of $\mathcal{G}$ under the historical measure $\P$. We would like to point out that, because of the L\'evy structure of the joint process $(S,N)$, the filtration $\G$ is right continous, so it satisfies the \emph{usual conditions}. In addition, we have the strict inclusion: 
\beqs
\Fc_n\subset \Gc_{\tau_n}\ \ \ {\rm for \ \  all}\  n \geq 1.
\enqs
We make an additional assumption on the model, which, among others, precludes arbitrage possibilities:

\noindent {\bf (NA)}
\beqs \int _0^T \frac{b^2(u)}{c^2(u)}du<\infty.
\enqs

\begin{Remark}\label{NA}
{\rm Underssumption {\bf (NA)}, we can define the probability measure $\Q$ by
\beqs
\frac{d\Q}{d\P}=e^{-\int _0^T \frac{b(u)}{c(u)}dW_u -\int _0^T \left (\frac{b(u)}{c(u)}\right)^2du}.
\enqs 
Under $\Q$, the process $N$ has the same law as under $\P$, and  $(\tau _n)_{n\geq 0}$ and $(S_t)_{0\leq t\leq T}$ are still independent under $\Q$.  Moreover, the process $S$ is a positive $(\Q,\G)$-local martingale so a supermartingale.  
This means that  the discrete-time process $(S_{\tau_n})_{n\geq 0}$ is a  $(\Q,\F)$ supermartingale as well.
}
\end{Remark}



\begin{Remark} \label{remcontX}
{\rm {\it Embedding in a continuous-time wealth process.} Given  $\alpha$ $\in$ $\Ac$ with corresponding wealth process $(X_{\tau_n})_n$ in \reff{selfX},  let us define  the continuous time process $(X_t)_{0\leq t<T}$ by
\beq \label{Xdis}
X_t &=& X_{\tau_{n}} + \alpha_{n} Z_{\tau_{n},t},  \;\;\; \tau_{n} < t \leq \tau_{n+1}, \;\; n \geq 0, \nonumber \\
&=& X_0 +\int _0^t H_udS_u, \;\;\;  0\leq t < T, \label{wealth-t} 
\enq
where $H$ is the simple and $\G$-predictable process
\beqs
H_t &=& \sum _{n=0}^{\infty} \frac{\alpha _n}{S_{\tau_{n}}} \mathbb{I}_{\{\tau _{n}<t\leq \tau_{n+1}\}},\ \ \ \ 0\leq t < T,
\enqs
representing the number of shares invested in the risky asset. 
 From  \reff{condadmiss} and since $S_t$ $>$ $0$, so $Z_{\tau_n,t}$ $>$ $-1$, $n$ $\geq$ $0$, we notice that the continuous time process $X$ is strictly positive: $X_t$ $>$ $0$ for  $0\leq t< T$. Moreover,  according to Remark \ref{NA},  $(X_t)_{0\leq t<T}$ is  a $(\Q,\G)$-local martingale, hence a super-martingale up to $T$. Consequently, we also have $X_{t-}>0$ for $0\leq t<T$.
 We would like to point out that the definition of $X_{\tau _n}$ in \reff{selfX} is consistent with \reff{wealth-t}, so $(X_{\tau_n})_{n\geq 0}$ is a positive $\F$-supermartingale under $\Q$.
Therefore, for each $\alpha \in \mathcal{A}$ we  may define  the terminal wealth value by:
\beqs
X_T \;  = \;  \lim_{n\rightarrow\infty} X_{\tau_n} \; = \; \lim_{t\nearrow T} X_t &=&  X_0 + \sum_{n=0}^\infty \alpha_n Z_{n+1}, 
\enqs
and, since $S_T=S_{T-}$ we also have
\beqs 
X_T 
&=& X_0 +\int _0^T H_udS_u,
\enqs
where the integrand $H$ is related to the simple trading strategy $\alpha$ as described above.
The supermartingale property implies  the budget constraint 
\beqs \mathbb{E}^{\Q}[X_T]\leq X_0.
\enqs
The continuous time wealth process $X$ has the meaning of a shadow wealth process: it is not observed except for at times $\tau _n$, $n\geq 0$. The no-short sale constraints \reff{condadmiss} is translated in terms of the number of shares held as
\beq \label{noshortcont} 
0 \; \leq H_t S_{t-} & \leq & X_{t-}, \;\;\; 0 \leq t < T. 
\enq
}
\end{Remark}

 We denote by $\Xc$ the set of all  
wealth processes $(X_t)_{0\leq t\leq T}$ given by \reff{Xdis}, by  using simple trading strategies under the no-short sale constraint \reff{condadmiss}/\reff{noshortcont}. 
We denote by $\bar\Xc$ the set of all positive wealth processes  $(X_t)_{0\leq t\leq T}$ given by \reff{Xdis}, by using \emph{general} $\G$-predictable and $S$-integrable processes $H$ satisfying \reff{noshortcont}.
We clearly have $\Xc$ $\subset$ $\bar\Xc$. 

For technical reasons, it is  sometimes convenient to regard trading strategies  equivalently in terms of proportions of wealth.  For any continuous time wealth process $X$ $\in$ $\bar\Xc$ associated to a trading strategy $H$ satisfying \reff{noshortcont}, let us consider the process $\pi$ $=$ $(\pi_t)_{0\leq t\leq T}$, defined by: $\pi_t$ $=$ $H_t S_{t-}/X_{t-}$,  and notice that $\pi$ is valued in $[0,1]$ 
by  \reff{noshortcont}.  We stress the dependence of the wealth on the proportion $\pi$, and denote by $X^{(\pi)}$ $=$ $X$, which is then written 
in a multiplicative way as
\beq \label{pi}
X_.^{(\pi)} = X_0\Ec\Big( \int_0^.  \pi \frac{dS}{S_{^-}}  \Big)  =  X_0 \Ec\Big( \int_0^.  \pi dL  \Big), 
\enq
 where $\Ec$ is the Dol\'eans-Dade operator.  Denote by $\Dc(\G)$ the set of all $\G$-predictable processes $\pi$ 
 valued in $[0,1]$. It is then clear that 
 \beqs \bar \Xc=\{X^{(\pi)}|\ \ \pi \in \Dc(\G)\}.
 \enqs

\section{Optimal investment problem and  dynamic programming}

\setcounter{equation}{0} \setcounter{Assumption}{0}
\setcounter{Theorem}{0} \setcounter{Proposition}{0}
\setcounter{Corollary}{0} \setcounter{Lemma}{0}
\setcounter{Definition}{0} \setcounter{Remark}{0}

 We investigate an optimal investment problem in the illiquid market described in the pre\-vious section.  Let us consider an utility function  
 $U$ defined on $(0,\infty)$,  strictly increasing, strictly  concave and $C^1$ on $(0,\infty)$, and satisfying the Inada conditions: $U'(0^+)$ $=$ $\infty$, 
 $U'(\infty)$ $=$ $0$. We make the following additional assumptions on the utility function $U$: 

\vspace{1mm}

\noindent {\bf (HU)} (i) there exist some constants $C>0$ and $p \in (0,1)$ such that 
\beqs
U^+(x)\leq C (1+x^p), \ \ (\forall)\ x>0
 \enqs
where $U^+$ $=$ $\max(U,0)$

\noindent (ii) Either $U(0)$ $>$ $-\infty$, 
  or $U(0)$ $=$ $-\infty$  and there exist some  constants $C'$ $>$ $0$ and  $p'$ $<$ $0$ such that
\beqs
U^-(x) & \leq & C(1+x^{p'}), \;\;\;  (\forall)\  x > 0,
\enqs
where $U^-$ $=$ $\max(-U,0)$.

The above assumptions include most popular utility functions, in particular those with constant relative risk aversion $1-p$ $>$ $0$,  
in the form $U(x)$ $=$ $(x^p-1)/p$, $x$ $>$ $0$.
 
Given the chosen positive  initial wealth $X_0$ $>$ $0$,  we  consider the optimal investment problem: 
\beq \label{defV0}
V_0 &=& \sup_{\alpha\in\Ac} \E[ U(X_T)]= \sup_{X\in \Xc} \E[ U(X_T)]. \enq
  Our aim is to provide an analytic solution to the control problem \reff{defV0} using direct dynamic programming, i.e. \emph{first} solve the Dynamic Programming Equation (DPE)  \emph{analytically} and \emph{then} perform a \emph{verification} argument. Therefore, there is no need to either  define the value function at later times or to prove the Dynamic Programming Principle (DPP).
  
 The Lemma below provides the intuition behind  the (DPE):
 \begin{Lemma}\label{conditional-expectation} Assume {\bf (HL)} holds true. 
 Let $\alpha \in \Ac$ and let $(X_{\tau _n})_{n\geq 0}$ be the wealth process associated with the trading strategy $\alpha$. Consider a measurable function
 $v:[0,T)\times (0, \infty)\rightarrow \R$. For a fixed $n\geq 0$, if $v(\tau _{n+1}, X_{\tau _{n+1}})\in L^1 (\Omega, \Gc, \P)$, then
 \beq
\E\big[ v(\tau_{n+1},X_{\tau_{n+1}}) | \Fc_n\big] 
&=& \int_{\tau_n}^T \int_{(-1,\infty)}  \lambda(s) e^{-\int_{\tau_n}^s \lambda(u) du} v(s,X_{\tau_n} + \alpha_{n} z) p(\tau_n,s,dz) ds, \nonumber \enq
where the above equality holds $\P$-a.s.
  \end{Lemma} 
{\bf Proof.}   From Remark \ref{rem:conditional-distribution} we have 
\beqs
\E\big[ v(\tau_{n+1},X_{\tau_{n+1}}) | \Fc_n\big] &=& \E\Big[ \E\big[ v(\tau_{n+1},X_{\tau_n} + \alpha_{n} Z_{n+1})  
| \Fc_n \vee \sigma( \tau_{n+1}) \big] 
\Big| \Fc_{n}  \Big] \nonumber \\
&=& \E\Big[   \int_{(-1,\infty)}  v(\tau_{n+1}, X_{\tau_n} + \alpha_{n} z ) p(\tau_n,\tau_{n+1},dz)  \Big | \Fc_n \Big]  \nonumber \\
&=& \int_{\tau_n}^T \int_{(-1,\infty)}  \lambda(s) e^{-\int_{\tau_n}^s \lambda(u) du} v(s,X_{\tau_n} + \alpha_{n} z) p(\tau_n,s,dz) ds. \nonumber 
\enqs \ep

\vspace{1mm}

Taking into account the above Lemma, we can now \emph{formally} write down the Dynamic Programming Equation as  
\beq \label{v}
v(t,x) &=&  \sup_{ a \in [0,x]} \int_t^T  \int_{(-1,\infty)}  \lambda(s) e^{-\int_t^s \lambda(u) du}  v(s,x+ az)p(t,s,dz) ds,   \nonumber \\
&=&  \sup_{ \pi \in [0,1]} \int_t^T  \int_{(-1,\infty)}  \lambda(s) e^{-\int_t^s \lambda(u) du}  v(s,x(1+ \pi z))p(t,s,dz) ds,  \label{fixedpoint} 
\enq 
for all $(t,x) \in [0,T)\times (0,\infty)$  
together with the natural  terminal condition
\beq\label{term}
\lim_{t\nearrow T, x'\rightarrow x} v(t,x') &=& U(x), \;\;\; x >0. 
\enq 
It appears that the right space of functions to be looking for a solution of 
\reff{v}-\reff{term} is actually the space
 $\Cc_U([0,T)\times (0,\infty))$ of measurable functions $w$ on $[0,T)\times (0,\infty)$, 
 such that 
\begin{enumerate}
\item $w(t,.)$ is concave on $(0,\infty)$ for all $t$ $\in$ $[0,T)$, and 
\item   for some  $C=C(w)$ $>$ $0$, we have 
\beq
U(x)  \; \leq \; w(t,x) & \leq & C(1+x), \;\;\; \forall (t,x) \in [0,T)\times (0,\infty) \label{wgrowth}.
\enq
\end{enumerate}
For any  $w$ $\in$ $\Cc_U([0,T)\times (0,\infty))$,  we consider the measurable function $\Lc w$  on $[0,T)\times (0,\infty)$ defined by:
\beq \label{defLc}
\Lc w(t,x)  &=&  \sup_{\pi\in [0,1]} \int_t^T  \int_{(-1,\infty)}  \lambda(s) e^{-\int_t^s \lambda(u) du}  w(s,x(1+ \pi z))p(t,s,dz) ds. 
\enq
Lemma \ref{operatorL} below shows that  the operator
$$\Lc:\Cc_U([0,T)\times (0,\infty))\rightarrow \Cc_U([0,T)\times (0,\infty)),$$ 
is well defined. Therefore, we are looking for a 
a solution $w$ $\in$ 
 $\Cc_U([0,T)\times (0,\infty))$ to the  DPE:   
 \beq \label{fixedpointLc}
\left \{
\begin{array}{ll}
 \Lc w = w\\
\lim _{t\nearrow T, x'\rightarrow x} w(t,x')=U(x).\end{array}\right.
 \enq
In order to solve the DPE and perform the verification argument, we need some technical details collected in the subsection below:

\subsection{A supersolution of the DPE and other technical details}  

\begin{Lemma}\label{operatorL} Assume that {\bf (HL)} holds. 
For any  $w$ $\in$ $\Cc_U([0,T)\times (0,\infty))$, $\Lc w$ also belongs to  $\Cc_U([0,T)\times (0,\infty))$.
For each $(t,x)\in [0,T)\times (0,\infty)$ the supremum in \reff{defLc} is attained at some $\pi (t,x)$ which can be chosen  measurable in $(t,x)$.

\end{Lemma}
{\bf Proof.} 
Given  $w$ $\in$ $\Cc_U([0,T)\times (0,\infty))$,  let us  consider the measurable function $\widehat w$ defined on  
$[0,T)\times (0,\infty)\times [0,1]$ by:
\beqs
\widehat w(t,x,\pi) &=&  \int_t^T  \int_{(-1,\infty)}  \lambda(s) e^{-\int_t^s \lambda(u) du}  w(s,x(1+ \pi z))p(t,s,dz) ds, 
\enqs
so that $\Lc w(t,x)=\sup _{\pi \in [0,1]} \widehat w(t,x,\pi)$.
Observe from \reff{wgrowth} and the integrability condition in Remark \ref{HZ} (ii)  that $\hat w$ is well-defined on $[0,T)\times (0,\infty)\times [0,1]$ and satisfies:
\beq \label{hatwgrowth}
-\infty \leq  \widehat w(t,x,\pi) & \leq & C(1+x), \;\;\; \forall (t,x,\pi) \in  [0,T)\times (0,\infty)\times [0,1],
\enq 
for some positive constant $C$ $>$ $0$ and, by \reff{wgrowth}, for  each  $(t,x)$ $\in$ $[0,T)\times (0,\infty)$,  we have
\beq \label{hatwU}
\widehat w(t,x,0) &\geq&  U(x).
\enq
 As a matter of fact, one can easily see that $\widehat w(t,x,\pi)$ is actually finite for any  $\pi \in [0,1)$ and may only equal negative infinity for $\pi=1$.
Consequently, for fixed $(t,x)$ $\in$ $[0,T)\times (0,\infty)$, $\widehat w(t,x,.)$ is a proper one-dimensional concave function defined on $[0,1]$ (concavity  follows easily from that of $w$). 
In addition, using the linear growth   \reff{wgrowth} together with Fatou lemma, we obtain that 
$\pi \in [0,1]\rightarrow \widehat w(t,x,\pi)$ is upper semicontinuous (this refers to the endpoints $\pi=0,1$ since the function is continuous on $(0,1)$ being finite and concave). Therefore
$\Lc w(t,x) = \max_{\pi\in [0,1]} \widehat w(t,x,\pi)$, where the maximum  is attained at some $\pi =\pi (t,x)$ which can be chosen measurable in $(t,x)$, 
see e.g. Ch. 11 in \cite{bershr78}.  In addition, since $\pi \rightarrow \widehat w(t,x,\pi)$ is continuous on $(0,1)$, the function $\Lc w$ has the additional representation
\beqs
\Lc w(t,x) &=& \sup_{\pi\in [0,1] \cap \mathbb{Q}} \widehat w(t,x,\pi),
\enqs 
which shows that $\Lc w$ is measurable.
The concavity of $w(t,.)$  implies the concavity of 
$(x,a)$ $\in$ $\{(x,a)\in (0,\infty)\times \R: a \in [0,x]\}$ $\rightarrow$ $\widehat w(t,x,a/x)$ for all $t$ $\in$ $[0,T)$. This  easily implies that $\Lc w(t,.)$ is also concave on $(0,\infty)$ for all $t$ $\in$ $[0,T)$.  Finally, it is  clear from \reff{hatwgrowth} and \reff{hatwU}  that $\Lc w$ satisfies also  the  growth condition: 
\beqs
 U(x)  \; \leq \; \Lc w(t,x) & \leq & C(1+x), \;\;\; \forall (t,x) \in [0,T)\times (0,\infty).  
\enqs
\ep

 The next lemma constucts a supersolution $f\in C_U([0,T)\times (0,\infty))$ for the DPE: 
\begin{Lemma} \label{supersolution}
Assume that  ${\bf (HL)}$,  ${\bf (NA)}$ and ${\bf (HU)}$   hold. 
Define 
\beqs
f(t,x)=\inf _{y>0}\left \{\E[\tilde U(yY_{t,T})] + y x \right \} \ \  (t,x)\in [0,T]\times (0,\infty),\ \enqs
 where 
\beq \label{YtT} Y_{t,T}=e^{-\int _t^T \frac{b(u)}{c(u)}dW_u -\int _t^T \left (\frac{b(u)}{c(u)}\right)^2du},\enq
and 
$\tilde U$ is the Fenchel-Legendre transform of $U$: 
 \beq \label{deftildeU}
 \tilde U(y) &=& \sup_{x>0} [U(x) - xy ] \; < \; \infty, \;\;\; \forall y > 0. 
 \enq
Then,  $f$ lies in the set $\Cc_U([0,T)\times (0,\infty))$, and satisfies
 \beq 
\left \{
\begin{array}{ll}
 \Lc f \leq  f\\
\lim _{t\nearrow T, x'\rightarrow x} f(t,x')=U(x).\end{array}\right.
 \enq
\end{Lemma}  
{\bf Proof.}
Jensen's inequality gives $\E[\tilde U(yY_{t,T})] \geq \tilde U(y),$ so
\beqs f(t,x)\geq \inf _{y>0}\left \{\tilde U(y) + y x\right\}=U(x).\enqs
From the  definition of $f$  we know that
\beq\label{upperbound} f(t,x)\leq  \E[\tilde U(yY_{t,T})] + y x,\ \ (\forall) \ y>0.
\enq
Fix a $y_0>0$. Jensen's inequality together with Assumption {\bf (HU)}(i) shows that
\beqs \mathbb{E}[\tilde U (y_0Y_{t,T})]\leq \mathbb{E}[\tilde U (y_0Y_{0,T})] <\infty,
\enqs
 so
 \beqs  f(t,x)\leq  \E[\tilde U(y_0Y_{t,T})] + y_0 x \leq C(1+x)\ \ \ (\forall)\ (t,x)\in [0,T]\times (0,\infty).
\enqs
This shows that $f$ $\in$ $\Cc_U([0,T)\times (0,\infty))$. 
Using assumption {\bf (HU)} (both (i) and (ii)) and the expression \reff{YtT}  we see that
\beqs
\lim _{t\nearrow T}\E[\tilde{U}(yY_{t,T})]=\tilde{U}(y), \ (\forall)\ y>0.
\enqs
We can now use this in \reff{upperbound} to deduce  that
\beqs
U(x)\leq \liminf _{t \nearrow T, x'\rightarrow x}f(t,x')\leq
\limsup _{t \nearrow T, x'\rightarrow x}f(t,x')\leq \tilde{U}(y)+xy, \ (\forall)\ y>0.
\enqs 
 Taking the infimum over $y$ we obtain the terminal condition. For each  fixed $t$, the function  $f(t,\cdot)$ is finite and concave on $(0,\infty)$, so the only thing left to check is the supersolution property.
Fix $0\leq t\leq s\leq T$ and $x>0$.
Denote by $h(z)=\mathbb{E}[\tilde U (z Y_{s,T})]$ and fix $y>0$ and $\pi \in [0,1]$.  By the very definition of the function $f$ we have  that
\beqs
f(s, x(1+\pi Z_{t,s}))\leq  h(yY_{t,s})+x(1+\pi Z_{t,s})yY_{t,s}.
\enqs
Using independence and the definition of $h$, we obtain 
\beqs \mathbb{E}[f(s, x(1+\pi Z_{t,s}))]&\leq& \mathbb{E} [ h(y Y_{t,s})+x(1+\pi Z_{t,s})yY_{t,s} ]= \\
& =&\mathbb{E}[\tilde U(yY_{t,s}Y_{s,T})]+\mathbb{E}[x(1+\pi Z_{t,s})yY_{t,s}] \leq \mathbb{E}[\tilde U(yY_{t,T})]+xy.
\enqs
Taking the inf over all $y$ and recalling the definition of $f(t,x)$ we obtain
\beqs f(t,x)\geq  \mathbb{E}[f(s, x(1+\pi Z_{t,s}))]= 
\int_{(-1,\infty)}  f(s,x(1+ \pi z))p(t,s,dz)
\enqs
for all $\pi$ and $s$. For a fixed $\pi$, we can integrate over $s$ to obtain
\beqs f(t,x)\geq  \int_t^T  \int_{(-1,\infty)} \lambda(s) e^{-\int_t^s \lambda(u) du} f(s,x(1+ \pi z))p(t,s,dz) ds,
\enqs
and then taking the supremum over $\pi$ we obtain
$f(t,x)\geq (\Lc f)(t,x)$, so the proof is complete. We would like to point out that, due to the linear growth condition on $f$, as well as Remark \ref{HZ} (ii), all expectations/integrals above are well defined, but may be negative infinity. In other words, the positive parts in all expectations/integrals  are actually integrable.
\ep
\begin{Remark}{\rm We would like to point out that the whole  analysis in this paper extends to the case when the Brownian part of the process $L$ is degenerate, as long as the jumps have full support on $(-1,\infty)$ and the jump measure allows for a martingale measure  with density process $Y$  that can replace the definition \reff{YtT} in the corresponding proofs.  In other words, the assumptions  {\bf (HL)}  and {\bf (NA)}  can be relaxed to include the situation when the drift can be removed by changing the jump measure appropriately, if the Gaussian part is missing.}
\end{Remark}

 It turns out that, for the verification arguments below, we also need an assumption on the integrability of jumps.

\noindent {\bf (HI)}:  
(i)  there exists $q>1$ such that
\beqs
 \int _0^T \int _{0}^{\infty}\Big( (1+ y)^{q}-1-q y \Big ) \nu (dt,dy)<\infty.
\enqs
(ii) If the utility function $U$ satisfies $U(0)=-\infty$ , then there exists $r<p'<0$ (where $p'$ is given in {\bf (HU)}(ii)) such that
\beqs
   \int _0^T \int _{-1}^{0}\Big ( (1+y)^{r}-1-ry \Big) \nu (dt,dy)<\infty.
\enqs

\noindent (iii) there are no predictable jumps, i.e. $\nu (\{t\}, (-1,\infty))=0$ for each $t$

\begin{Remark}{\rm Using convexity, it is an easy exercise to see that assumption {\bf (HI)} (i) 
can actually be rephrased as $\nu_q([0,T]) <\infty$, and assumption {\bf (HI)}(ii) as
$\nu_r([0,T])<\infty$
where
\beqs
\nu_l(dt)= \int _{-1}^{\infty}\sup _{\pi \in [0,1]}\left ( (1+\pi y)^{l}-1-l\pi y \right ) \nu(dt,dy).
\enqs

}
\end{Remark}
We now prove a crucial  uniform integrability condition, but before that we denote  by $\Tc$ the set of random times $0\leq \tau <T$ which are stopping times with respect to the filtration $\G $.
\begin{Lemma} \label{lemuniU}
Assume that {\bf (HL)}, {\bf (HU)}, {\bf (NA)} and {\bf (HI)}  hold.

\noindent {\bf (1)} For any  $X$ $\in$ $\bar\Xc$, the family $(f^+(\tau , X_{\tau }))_{\tau \in  \Tc}$  is uniformly $\P$-integrable.  
  
\noindent {\bf (2)} For any $X$ $\in$ $\bar\Xc$, the family $(U^-(X_{\tau}))_{\tau \in \Tc}$, is uniformly $\P$-integrable.

\end{Lemma}
{\bf Proof.} Assume $\nu _l([0,T])<\infty$ for some $l$.
 Consider $X$ $=$ $X^{(\pi)}$ $\in$ $\bar\Xc$ for some $\pi$ $\in$ 
$\Dc(\G)$, and recall that 
\beq \label{expressXpi}
dX^{(\pi)}_t &=&\pi _t X^{(\pi)}_ {t-}dL_t,\ \ 0\leq t\leq T.
\enq
In order to simplify notation, we supress the upper indices of $X$. We apply It\^o formula to $(X_t)^l$ to conclude that
\beqs
X_t^l=x^l+\int _0^t (X_{u-})^l \Big(l \pi _u b(u)+l(l-1)c^2(u)\pi ^2(u)\Big ) du+\\
\int _0^t\int _{-1}^{\infty}(X_{u-})^l \Big ((1+\pi_u y)^l-1-l\pi _uy\Big )\nu (du,dy)+''local \ martingale''
\enqs
Fix a  stopping  time $\tau \in \Tc $. If $T'_n$ is a sequence of localizing stopping times for the local martingale part, denote by 
 $T_n = T'_n \wedge \big\{ \inf t : (X_{t})^{l} \geq n \big\}$. 
 Observe that $T_n \nearrow T$ a.s. since 
 $X_{-}$ is locally bounded and locally bounded away from zero. We then have, for each $0\leq t<T$,
\beq 
\E[(X_{t\wedge \tau  \wedge T_n})^l] &=& x^l+  \E \Big[\int _0^{t\wedge \tau \wedge T_n} (X_{u-})^l\Big \{\Big(l \pi _u b(u)+l(l-1)c^2(u)\pi ^2(u)\Big )du  \nonumber
\\
& &  \;\;\; + \;  \int _{-1}^{\infty}\Big ((1+\pi_u y)^l-1-l\pi _uy\Big )\nu (du,dy)\Big \} \Big] \nonumber \\
\leq  \;\; x^l &+& \E\left [\int _0^{t\wedge \tau \wedge T_n}(X_{u-})^l \Big \{\Big (|l b(u)|+|l(l-1)c^2(u)|\Big )du+ \nu _l(du)\Big \}\right ] \label{ineq-before-gronwall}
\enq
Since $(X_{u-})^l\leq n$ for $0\leq u\leq \tau \wedge T_n$ and $\nu _l([0,T])<\infty$, we conclude that 
\beq \label{finite-stopped}
\E[(X_{t\wedge \tau  \wedge T_n})^l]<\infty,\ \ \ 0\leq t< T.
\enq
In addition, since the paths of the process $X^l$ are RCLL and $\nu_l (\{u\})=0$ for each $0\leq u\leq T$ (because of {\bf (HI)} part (iii)), we have that, for each $0\leq t<T$,  with $\P$-probability one
\beqs
\int _0^{t\wedge \tau \wedge T_n}(X_{u-})^l \Big \{\Big (|l b(u)|+|l(l-1)c^2(u)|\Big )du+ \nu _l(du)\Big \} = 
\\
\int _0^{t\wedge \tau \wedge T_n}(X_u)^l \Big \{\Big (|l b(u)|+|l(l-1)c^2(u)|\Big )du+ \nu _l(du)\Big \} \leq \\
\int _0^t(X_{u \wedge \tau \wedge T_n})^l \Big \{\Big (|l b(u)|+|l(l-1)c^2(u)|\Big )du+ \nu _l(du)\Big \}. \enqs
Replacing this in \reff{ineq-before-gronwall} and using Fubini, we obtain
\beq \label{ineq-gronwall}
\E[(X_{t\wedge \tau  \wedge T_n})^l] \leq x^l+\int _0^t \E[(X_{u\wedge \tau  \wedge T_n})^l] \Big \{\Big (|l b(u)|+|l(l-1)c^2(u)|\Big )du+ \nu _l(du)\Big \}.\enq
Now,  using \reff{finite-stopped} and  $\nu _l ([0,T])<\infty$, we can apply Gronwall  in \reff{ineq-gronwall} to conclude that
$
\E[(X_{t\wedge \tau \wedge T_n })^l]\leq M(l)<\infty $,  for each $0\leq t< T$, where $M(l)$ does not depend on $\tau$ or $n$. Letting $n \rightarrow \infty$  and $t\rightarrow T$, by Fatou, we obtain  $\E[(X_{\tau})^l]\leq M(l)$
for each stopping time $\tau \in \mathcal{T}$.

We can finish the proof considering $l=q$ for item (i) and   $l=r$ for item (ii), and also using the upper bound $f(t,x)\leq C(1+x)$ as well as Assumption {\bf (HU)} part (ii).
\ep

\begin{Remark}\label{finite}{\rm  In case $U(0)=-\infty$, we can follow the arguments in the Proof of Lemma \ref{lemuniU} for the case $\pi=1$ and $l=r$ (taking into account that $X_t=X_0S_t$ for $0\leq t<T$) to conclude that 
\beqs \mathbb{E}[ (S_t)^r]= \mathbb{E}[ (1+Z_{0,t})^r]\leq C(0)<\infty \ {\rm for}\ 0\leq t<T.
\enqs 
(we assumed that $S_0=1$ above, and we also used that the times $0\leq t<T$, because are deterministic, belong to $\Tc$). The same argument actually works if we start at any time $0\leq t<T$, so we have 
\beqs
\mathbb{E}[ (1+Z_{t,s})^r]=\int _{(-1,\infty)} (1+z)^r p(t,s,dz)\leq C(t)<\infty, \ {\rm for}\ \ t\leq s <T.
\enqs
}
\end{Remark}

\subsection{Construction of a solution for the DPE}

We provide a constructive  proof for the existence of a solution of  \reff{fixedpointLc} using an iteration scheme.  Let us define inductively the sequence of functions 
$(v_m)_m$ in $\Cc_U([0,T)\times (0,\infty))$ by: 
\beqs
v_0 &=& U, \;\;\; v_{m+1} \; = \; \Lc v_m, \;\;\; m \geq 0. 
\enqs
 \begin{Lemma} Assume {\bf (HL)}, {\bf (NA)}, and {\bf (HU)}. Then the sequence of functions $v_m$ satisfies
 \beqs
 v_m\leq v_{m+1}\leq f,  \;\;\; m \geq 0. \enqs
 \end{Lemma}
 {\bf Proof.} We do the  proof by induction. We obviously have $U=v_0\leq v_1$. In addition, since the operator $\Lc$ is monotone and $U\leq f$ we have 
 \beqs v_1=\Lc U\leq \Lc f \leq f,
 \enqs
 so the statement is true for $m=0$. Assume now the statement is true for $m$.
 We use again the monotonicity of $\Lc$ to get
 \beqs v_{m+2}=\Lc v_{m+1}\geq \Lc v_m=v_{m+1},\ \ \ \ \ \ v_{m+2}=\Lc v_{m+1}\leq \Lc f\leq f,
 \enqs
 so the proof is finished.
 \ep
 
 \vspace{2mm}
 
 Under the conditions of the above Lemma, the nondecreasing sequence $(v_m)_m$ converges pointwise, and we may define
\beq \label{defv*}
v^* &=& \lim_{m\rightarrow\infty} v_m \; \leq \; f. 
\enq
We   show  next that $v^*$ satisfies the fixed point DP equation.

\vspace{1mm}

 \begin{Theorem} \label{theofixedpoint}
 Assume that {\bf (HL)}, {\bf (NA)}, {\bf (HU)} and  {\bf (HI)}   hold.  Then,   
 $v^*$ is solution to  the fixed point DP  (\ref{fixedpointLc}).  
\end{Theorem}
{\bf Proof.}   Fix  $\pi \in [0,1]$. We know by construction that 
\beqs 
v_{m+1}(t,x)  &\geq &  \int_t^T  \int_{(-1,\infty)}  \lambda(s) e^{-\int_t^s \lambda(u) du}  v_m(s,x(1+ \pi z))p(t,s,dz) ds. 
\enqs
If $0\leq \pi <1$, then $v_m(s, x(1+\pi z)\geq U( x(1-\pi))$ so the integral on the right hand side is clearly finite.
If $\pi=1$, according to Remark \ref{finite}, the integral on the right hand side is still finite for each $m\geq 0$. Therefore, we can 
let $m\ \nearrow \infty$ and use the  monotone convergence theorem to  obtain
\beqs 
v^*(t,x)  &\geq &  \int_t^T  \int_{(-1,\infty)}  \lambda(s) e^{-\int_t^s \lambda(u) du}  v^*(s,x(1+ \pi z))p(t,s,dz) ds. 
\enqs
Since this happens for each $\pi$, taking the supremum over $\pi$ we get
$v^*\geq \mathcal{L}v^*$. Conversely, for $\varepsilon >0$ there exists $m$ such that
$v^*(t,x)-\varepsilon \leq v_{m+1}(t,x)$ and (because of convexity the maximum is attained) $\pi^m(t,x)$ such that
\beqs 
v_{m+1}(t,x)  &= &  \int_t^T  \int_{(-1,\infty)}  \lambda(s) e^{-\int_t^s \lambda(u) du}  v_m(s,x(1+ \pi^m(t,x) z))p(t,s,dz) ds. 
\enqs
Since $v_m \leq v^*$ it follows that
\beqs 
v^*(t,x)-\varepsilon & \leq &    \int_t^T  \int_{(-1,\infty)}  \lambda(s) e^{-\int_t^s \lambda(u) du}  v^*(s,x(1+ \pi^m(t,x) z))p(t,s,dz) ds \\
&\leq & \mathcal{L}v^* (t,x). 
\enqs
Letting $\varepsilon \rightarrow 0$ we obtain $v^*=\mathcal{L} v^*$. Finally, since $U(t,x)\leq v^*(t,x)\leq f(t,x)$ and the  function $f$ satisfies the boundary condition (\ref{term}) by Lemma \ref{supersolution},  we conclude that  $v^*$ is a solution to the fixed point DP equation \reff{fixedpointLc}.   
\ep

\begin{Remark}
{\rm The previous theorem shows the existence of a fixed point to the DP equation \reff{fixedpointLc}, and gives also an iterative procedure for constructing 
a fixed point. In the next subsection, we shall prove that such a fixed point is equal to the value function $v$, which implies in particular 
the uniqueness for the fixed point equation \reff{fixedpointLc}. 
}
\end{Remark}

\subsection{Verification and  optimal strategies}
Consider the solution $v^*$  to the fixed point DP equation \reff{fixedpointLc}, constructed in Theorem \ref{theofixedpoint}.  We now state a verification theorem for the fixed point equation \reff{fixedpointLc}, which provides the optimal portfolio strategy in feedback form.

\begin{Theorem} \label{theoverif}
Assume that {\bf (HL)}, {\bf (NA)},  {\bf (HU)} and {\bf (HI)}  hold. Then,
\beqs V_0=v^*(0,X_0),
\enqs and an optimal control  $\hat\alpha$ $\in$  $\Ac$   is given by
\beq \label{defhatalpha}
\hat\alpha_{n} &=& \hat\pi(\tau_n,\hat X_{\tau_n}) \hat X_{\tau_n}, \;\;\; n \geq 0, 
\enq
where $\hat\pi$ is a measurable function on $[0,T)\times (0,\infty)$ solution to
\beqs
\hat\pi(t,x) & \in & {\rm arg}\max_{\pi\in [0,1]}  \int_t^T  \int_{(-1,\infty)}  \lambda(s) e^{-\int_t^s \lambda(u) du}  v^*(s,x(1+ \pi z))p(t,s,dz) ds,
\enqs
and $(\hat X_{\tau_n})_{n\geq 0}$ is the wealth given by 
\beqs
\hat X_{\tau_{n+1}} &=& \hat X_{\tau_n}+ \hat\alpha_{n} Z_{n+1}, \;\;\; n \geq 0, 
\enqs
and starting from  $\hat X_0$ $=$ $X_0$. 
\end{Theorem} 
{\bf Proof.}  
Consider $\alpha$ $\in$ $\Ac$ and the corresponding positive wealth process $(X_{\tau_n})_{n\geq 0}$.  From Lemma \ref{lemuniU}, we know that 
\beqs
\mathbb{E}[|v^*(\tau _n, X_{\tau _n})|]<\infty,\ \ \ \ \ (\forall)\ n\geq 0.
\enqs
We apply Lemma \ref{conditional-expectation} to get for any $n$ $\geq$ $0$:
\beq
\E\big[ v^*(\tau_{n+1},X_{\tau_{n+1}}) | \Fc_n\big] &=& \int_{\tau_n}^T \int_{(-1,\infty)}  \lambda(s) e^{-\int_{\tau_n}^s \lambda(u) du} v^*(s,X_{\tau_n} + \alpha_{n} z) p(\tau_n,s,dz) ds \nonumber \\
& \leq & \Lc v^*(\tau_n,X_{\tau_n}) \; = \; v^*(\tau_n, X_{\tau_n}), \label{inegvsur}
\enq
so  the process  $\{v^*(\tau_n,X_{\tau_n}),n \geq 0\}$ is a 
$(\P,\F)$-supermartingale. Recalling that $v^*(t,.)$ $\geq$ $U$ we obtain
\beqs
\E[U(X_{\tau_n})]  \;  \leq \; \E[ v^*(\tau_n,X_{\tau_n})] & \leq & v^*(0,X_0), \;\;\;\;\; (\forall) \; n \geq 0. 
\enqs
Now, by Lemma \ref{lemuniU}, the sequence $(U(X_{\tau_n}))_n$ is uniformly integrable. By sending $n$ to infinity into the last inequality, we 
then get
\beqs
\E[U(X_T)] & \leq & v^*(0, X_0).
\enqs
Since $\alpha$ is arbitrary, we obtain   $V_0$ $\leq$ $v^*(0,X_0)$. 

Conversely, let $\hat\alpha$ $\in$ $\Ac$ be the portfolio strategy given by \reff{defhatalpha}, and $(\hat X_{\tau_n})_{n\geq 0}$ the associated 
wealth process. Then, by the same calculations as in \reff{inegvsur}, we have now the equalities:
\beqs
\E\big[ v^*(\tau_{n+1},\hat X_{\tau_{n+1}}) | \Fc_n\big] &=& 
\int_{\tau_n}^T \int_{(-1,\infty)}  \lambda(s) e^{-\int_{\tau_n}^s \lambda(u) du} v^*(s,X_{\tau_n} + \hat \alpha_{n} z) p(\tau_n,s,dz) ds \\
&=& \Lc v^*(\tau_n,\hat X_{\tau_n}) \; = \; v^*(\tau_n,\hat X_{\tau_n}), \;\;\; n \geq 0, 
\enqs
by definition of $\Lc$ and $\hat\alpha$. This means that the process  $\{v^*(\tau_n,\hat X_{\tau_n}),n \geq 0\}$ is a $(\P,\F)$-martingale, and so:
\beqs
\E[ v^*(\tau_n,\hat X_{\tau_n})] & = & v^*(0,X_0), \;\;\;\;\;  (\forall) \;  n \geq 0. 
\enqs
From the the bounds $U\leq v^*\leq f$  and Lemma \ref{lemuniU}, we know  that the sequence 
$(v^*(\tau_n,\hat X_{\tau_n}))_n$ is uniformly integrable. By sending $n$ to infinity into the last equality, and recalling the terminal 
condition for $v^*$,  we then get
\beqs
\E[ U(\hat X_T)] & = & v^*(0,X_0). 
\enqs
Together with the inequality, $V_0$ $\leq$ $v^*(0,X_0)$, this proves that $V_0$ $=$ $v^*(0,X_0)$ and $\hat\alpha$ is an optimal control. 
\ep

\vspace{2mm}

An identical verification argument  to the proof of Theorem \ref{theoverif} can be performed for an investor starting at time $t$ with initial capital $x$: this way we prove that $v^*$ is actually the value function of the control problem. In addition, this  shows that the Dynamic Programming  Equation \reff{fixedpointLc} has a unique solution.  For the sake of avoiding the heavy notation associated with strategies starting at time $t$, we decided to only do the verification for time $t=0$.

 The Proposition below  shows  that actually we can approximate the optimal control, and not only the maximal expected utility,  using  a finite number of iterations. The approximate optimal control is actually very simple, since after the $m$-th  arrival time  all the wealth is invested in the money market. In addition, a stochastic control representation for the iteration $v_m$ is provided.

\begin{Proposition}\label{v-n}
Assume that  {\bf (HL)}, {\bf (NA)},  {\bf (HU)} and {\bf (HI)}  hold. Then
\beq \label{defv-n}
v_m(0,X_0) &=& \sup_{\alpha \in \Ac_m} \E[ U(X_T)], 
\enq
where $\Ac_m$ is the set of admissible controls $\alpha$ $=$ $(\alpha_n)_{\geq 0}$ $\in$ $\Ac$  
such that  all money is invested in the money market  after $m$ arrivals,  i.e. $\alpha_ n$ $=$ $0$ for  $n$ $\geq$ $m$.  

For any $0\leq n\leq m-1$, consider the measurable function $\hat \pi^n(\cdot, \cdot)$ defined by 
\beqs
\hat\pi^n (t,x)& = & {\rm arg}\max_{\pi\in [0,1]}  \int_t^T  \int_{(-1,\infty)}  \lambda(s) e^{-\int_t^s \lambda(u) du}  v_{m-n-1}(s,x(1+ \pi z))p(t,s,dz) ds,
\enqs
so that
\beqs
v_{m-n} (t,x) & = &  \int_t^T  \int_{(-1,\infty)}  \lambda(s) e^{-\int_t^s \lambda(u) du}  v_{m-n-1}(s,x(1+ \pi^n(t,x)  z))p(t,s,dz) ds.
\enqs
Define in feedback form the admissible strategy  $\hat\alpha^m$ $\in$ $\Ac_m$  by  
$\alpha^m_{n}$ $=$ $\hat\pi^n(\tau_n,\hat X_{\tau_n}^m)\hat X_{\tau_n}^m$ for  $0\leq n\leq m-1$ and $\alpha ^m_n=0$ for $n\geq m$,  where the wealth processes 
$(\hat X^m_{\tau_n})_{n\geq 0}$ is  given by 
\beqs
\hat X^m_{\tau_{n+1}} &=& \hat X^m_{\tau_n} + \hat\alpha_{n}^m Z_{n+1}, \;\;\; 0\leq n\leq m-1,\ \ \ \ \ \  
\hat X^m_{\tau_n} = \hat X^m_{\tau_m},\ \ \ n\geq m,
\enqs 
starting from the initial wealth $X_0$. Then $\alpha ^m\in \Ac_m$ is an optimal control for \reff{defv-n}.
\end{Proposition}
{\bf Proof.}   The proof is based on similar arguments to the proof of Theorem \ref{theoverif}. Namely, for each $\alpha \in \Ac_m$, one can use Lemma \ref{conditional-expectation} to conclude that
$(v_{m-n}(\tau _n, X_n))_{n=0,1, \dots, m}$ is a supermartingale and, for the particular choice of the control $\alpha ^m$ described above we actually have that 
$(v_{m-n}(\tau _n, \hat X^m_n))_{n=0,1, \dots,m}$ is a true martingale. Since $v_0(t,x)=U(x)$ and for each $\alpha \in \Ac_m$ the wealth process $X$ is constant after the arrival time $\tau _m$, it is easy to finish the proof.
\ep

\vspace{2mm}

Theorems  \ref{theofixedpoint} and \ref{theoverif} together  show how we can compute by iterations  the maximal expected utility and  the optimal control. Since the control problem is finite-horizon in time and infinite horizon in $n$, taking into account Proposition \ref{v-n}, the iteration procedure represents exactly the approximation of the infinite horizon problem by a sequence of finite horizon problems. 

\subsection{Example: the case of CRRA utility functions}

\setcounter{equation}{0} \setcounter{Assumption}{0}
\setcounter{Theorem}{0} \setcounter{Proposition}{0}
\setcounter{Corollary}{0} \setcounter{Lemma}{0}
\setcounter{Definition}{0} \setcounter{Remark}{0}

In this subsection, we consider the case of CRRA (Constant Relative Risk Aversion) utility functions.

First, let us take logarithm utility functions:  
\beqs
U(x) &=& \ln x, \;\;\; x > 0.
\enqs  
We easily see that in this case the value function has the form:
\beqs
v(t,x) &=& U(x) + \varphi(t),
\enqs 
 for some nonnegative continuous function $\varphi$ on $[0,T)$ with $\varphi(T^-)$ $=$ $0$.  The computation of  $\Lc v$ in \reff{defLc} is straightforward:
\beqs
\Lc v &=& \ln x  +  \int_t^T \lambda(s) e^{-\int_t^s \lambda(u) du} \varphi(s) ds  \\
& & \;\;\; + \;  \sup_{\pi\in [0,1]} \int_t^T   \lambda(s) e^{-\int_t^s \lambda(u) du} \Big( \int_{(-1,\infty)}  \ln(1+\pi z) p(t,s,dz) \Big) ds.  
\enqs 
Hence, we see that $v$ is a solution to the fixed point equation: $\Lc v$ $=$ $v$ iff  $\varphi$ satisfies:
\beqs 
\varphi(t) &=&  \int_t^T \lambda(s) e^{-\int_t^s \lambda(u) du} \varphi(s) ds +   F(t), \;\;\; t \in [0,T), 
\enqs
where $F$ is the function defined on $[0,T)$ by 
\beqs
F(t) &=& \sup_{\pi\in [0,1]} \int_t^T   \lambda(s) e^{-\int_t^s \lambda(u) du} \Big( \int_{(-1,\infty)}  \ln(1+\pi z) p(t,s,dz) \Big) ds. 
\enqs

 

\vspace{2mm}

We next consider power utility functions: 
\beqs
U(x) &=& \frac{x^\gamma}{\gamma}, \;\;\; x > 0, \; \gamma < 1,  \gamma \neq 0, \;\;
\enqs 
In this case, the value function has the form:
\beqs
v(t,x) &=&  \varphi(t) U(x),
\enqs 
 for some continuous function $\varphi$ on $[0,T)$, greater than $1$,  with $\varphi(T^-)$ $=$ $1$.  We easily compute $\Lc v$ in \reff{defLc}:
\beqs
\Lc v &=&  U(x)   \;  \sup_{\pi\in [0,1]} \int_t^T   \lambda(s) e^{-\int_t^s \lambda(u) du} \varphi(s) \Big( \int_{(-1,\infty)}  (1+\pi z)^\gamma p(t,s,dz) \Big) ds.  
\enqs 
Hence, we see that $v$ is a solution to the fixed point equation: $\Lc v$ $=$ $v$ iff  $\varphi$ satisfies:
\beqs 
\varphi(t) &=&  \sup_{\pi\in [0,1]} \int_t^T   \lambda(s) e^{-\int_t^s \lambda(u) du} \varphi(s) \Big( \int_{(-1,\infty)}  
(1+\pi z)^\gamma p(t,s,dz) \Big) ds, \;\; t \in [0,T). 
\enqs

\section{Convergence in  the  illiquid market model}

\setcounter{equation}{0} \setcounter{Assumption}{0}
\setcounter{Theorem}{0} \setcounter{Proposition}{0}
\setcounter{Corollary}{0} \setcounter{Lemma}{0}
\setcounter{Definition}{0} \setcounter{Remark}{0}

So far, we have considered the optimal investment problem \reff{defV0} for a \emph{fixed} arrival rate function 
 $\lambda :[0,T)\rightarrow [0,\infty)$ satisfying condition \reff{hypintens}. To emphasize the dependence on the arrival rate, let us denote by $V_0^{\lambda}$ the value  in \reff{defV0}.  When the arrival rate is very large \emph{at all times} (in some sense to be precised), 
 one would expect that $V_0^\lambda$  is very close to the optimal expected utility of an agent who can trade at all times (therefore continuously) in the asset $S$.  It is also expected that the constraint \reff{condadmiss}, which is {\it implicitly} contained in the admissibility condition \reff{xtaunpos} 
in the discrete time  illiquid case, becomes an {\it explicit} no-short constraint \reff{noshortcont} in the continuous time limit. 
This section is devoted to proving that this is actually true.
 
 First, we need to define the optimization problem for the agent who can trade continuously.
We remind the reader that continuous time trading strategies can be defined by \reff{Xdis}. 
We denote by $\Xc ^S$ the set of positive wealth processes  $(X_t)_{0\leq t\leq T}$ given by \reff{Xdis}, by using $\G ^S$-predictable and $S$-integrable processes $H$ satisfying the no-short sale constraint \reff{noshortcont}.
The filtration  $\G^S$ $=$ $(\Gc^S_t)_{0\leq t\leq T}$ is defined by 
\beqs 
\Gc_t^S=\sigma\{S_s, 0\leq  s\leq t\}\vee \Nc,
\enqs
and represents the information one can get from following the asset $S$. Because of the L\'evy structure of $S$, $\G^S$ satisfies the usual conditions.
We also   denote by $\Dc(S)$  the set of all $\G^S$-predictable processes $\pi$ 
 valued in $[0,1]$. It is then clear that 
\beqs  \Xc ^S=\{X^{(\pi)}|\ \ \pi \in \Dc(S)\}.
 \enqs
The optimization problem for an agent trading continuously, under no-short selling constraints can be formulated as
\beq \label{defvMer}
V^{_M}_0 &=&  \sup_{X \in \Xc^S} \E[U(X_T)]= \sup_{\pi  \in \Dc(S)} \E[U(X^{(\pi)}_T)].
\enq

 The main result of this section is:

\begin{Theorem}\label{asymptotic}
Under Assumptions {\bf (HL)}, {\bf (NA)},  {\bf (HU)} and {\bf (HI)},  consider $(\lambda _k)_k$ a sequence of intensity functions satisfying 
\reff{hypintens}.  If  
\beq \label{infintens}
\sum_{k=0}^{\infty} \exp\left( -\int _t^s \lambda _k (u)\,du\right)<\infty,\ \ (\forall)\ \ 0\leq t<s<T,
\enq
 then
\beqs V_0^{\lambda _k}\rightarrow V_0^M, \;\;\; \mbox{ as } k \; \mbox{ goes to infinity}, 
\enqs
where $V_0^M$ is defined by \reff{defvMer}.
\end{Theorem}

\begin{Remark}
{\rm Condition \reff{infintens}  is satisfied for example with $\lambda_k(t)$ $=$ $k . \lambda(t)$, where $\lambda$ is an intensity function satisfying \reff{hypintens}. This condition also implies that   for all  $0\leq t<s<T$, 
\beqs
\int_t^s \lambda_k(u) du & \rightarrow  & \infty,   \;\;\; \mbox{ as } k \; \mbox{ goes to infinity}. 
\enqs
}
\end{Remark}

\vspace{1mm}

In order to prove Theorem \ref{asymptotic}, we first have to put all the optimization problems \reff{defV0} on \emph{the same physical probability space}, independent of the intensity function $\lambda$. This is an easy task actually. We consider a probability space $(\Omega,\Gc,\P)$ supporting two independent processes: the continuous time stock price process $(S_t)_{0\leq t\leq T}$ (which has all the desired properties) and a Poisson process $(M_t)_{0\leq t <\infty}$ with intensity equal to one. After that, for each intensity function $\lambda$ we define the nonhomogenous Poisson process $N^{\lambda}$ (actually its sequence of jumps) by \reff{representation-N}. Therefore, for different intensities, we still have the same physical space.
We now  denote by $\mathbb{F}^{\lambda}$ and
 $\mathbb{G}^{\lambda}$ the discrete and continuous time filtrations on $(\Omega,\Gc,\P)$  defined by \reff{F} and \reff{cont-filtration} corresponding to the intensity $\lambda$, and by $\tau_n^\lambda$ the associated jump times.

The main obstacle in proving Theorem \ref{asymptotic} is the fact that the filtration $\mathbb{F}^{\lambda}$ only observes the process $S$ at the arrival times, while the filtration $\G^S$ used by the investor in \reff{defvMer} observes the stock continuously. This  problem is overcome in three steps

\vspace{2mm}

\noindent {\bf Step 1}:  first, we show that in \reff{defV0}, the discrete-time filtration $\mathbb{F}^\lambda =(\mathcal{F}^{\lambda}_n)_{n\geq 0}$ can be replaced by the larger filtration $(\mathcal{G}^{\lambda}_{\tau^{\lambda} _n})_{n\geq 0}$. In other words, due to the Markov structure of the model, an investor who can only trade at the discrete arrival times, cannot improve his/her expected utility by continuously observing the evolution of the stock  between the arrival times. This is done in Lemma  \ref{verification} below.

\begin{Lemma} \label{verification} Fix an intensity function $\lambda$ and define 
\beq \label{defV0c}
V^{\lambda ,c}_0 & := & \sup_{\alpha\in\Ac ^{\lambda} _c} \E[ U(X_T)],\enq
where $\Ac ^{\lambda} _c$ is the set of simple admissible strategies $\alpha =(\alpha _n)_{n\geq 0}$ with continuous observation, which means that for each $n\geq 0$ we have  $\alpha _n\in \mathcal{G}^{\lambda}_{\tau^{\lambda} _n}$ and $\alpha$ satisfies the constraint \reff{condadmiss} for  the wealth process $(X_{\tau _n})_{n\geq 0}$ defined by \reff{selfX}.

Then, under   Assumptions {\bf (HL)}, {\bf (NA)},  {\bf (HU)} and {\bf (HI)}, we have $V_0^{\lambda,c}=V_0^{\lambda}$.
\end{Lemma}
{\bf Proof.}
Similarly to Remark  \ref{rem:conditional-distribution},  Assumption ${\bf (HL)}$  together with the independence of $S$ and $N$ ensures that  for all $n$ $\geq$ $0$, 
the (regular) distribution of $(\tau _{n+1}, Z_{n+1})$ conditioned on $ \Gc ^{\lambda} _{\tau ^{\lambda}_{n}} $ is given by:
\begin{enumerate}
\item $
\P[ \tau^{\lambda}_{n+1} \in ds | \Gc ^{\lambda} _{\tau ^{\lambda}_{n}} ] =\lambda(s) e^{-\int_{\tau _n}^s \lambda(u) du}ds$
\item further conditioning on knowing  the next arrival time $\tau ^{\lambda}_{n+1}$, the return $Z_{n+1}$ has distribution
$$\P[Z_{n+1}\in dz| \Gc ^{\lambda} _{\tau ^{\lambda}_{n}}  \vee \sigma (\tau ^{\lambda} _{n+1})]=p(\tau ^{\lambda} _n, \tau ^{\lambda} _{n+1}, dz).$$
\end{enumerate}

Therefore, in Lemma \ref{conditional-expectation}, one can replace the filtration $\F^{\lambda}$ by the larger filtration $(\mathcal{G}^{\lambda}_{\tau^{\lambda} _n})_{n\geq 0}$ and obtain that, for each $\alpha \in \Ac _c^{\lambda}$ we have 
\beqs
\E\big[ v(\tau^{\lambda} _{n+1},X_{\tau ^{\lambda}_{n+1}}) |    \Gc ^{\lambda} _{\tau ^{\lambda}_{n}}        \big] 
&=& \int_{\tau ^{\lambda}_n}^T \int_{(-1,\infty)}  \lambda(s) e^{-\int_{\tau ^{\lambda}_n}^s \lambda(u) du} v(s,X_{\tau ^{\lambda}_n} + \alpha_{n} z) p(\tau ^{\lambda}_n,s,dz) ds.  
\enqs
After that, one can just follow
 the verification arguments in the  proof of Theorem \ref{theoverif} to show that
 \beqs
 V_0^{\lambda ,c}=v^{*,\lambda}(0,X_0),
 \enqs
 which ends the proof.
 \ep

\vspace{2mm}

\noindent {\bf Step 2}: we define the continuous time filtration $\mathbb{G}^{\infty}=(\mathcal{G}^{\infty}_t)_{0\leq t\leq T}$ which contains all the information from the arrival times right at time zero. This corresponds to an investor who knows in advance all the jumps of the homogeneous Poisson process $M$  and also observes continuously the stock $S$ up to time $t$:
\beqs 
\mathcal{G}^{\infty}_t=\mathcal{G}^S_t \vee \sigma (M_u:\ 0\leq u<\infty),\ \ \ 0\leq t<T.
\enqs
Because the information added is independent, the process $S$ is still a semimartingale with respect to the larger filtration $\mathbb{G}^{\infty}$, which satisfies the usual conditions as well.  Using again the independence property,  Lemma \ref{lemmer} below shows that if the investor in \reff{defvMer} has the additional information in $\G ^{\infty}$,  he/she cannot improve the maximal expected utility.

\begin{Lemma} \label{lemmer} Consider the set $\bar \Xc _c$ of wealth processes defined by \reff{wealth-t} where the general integrand $H$ is $\G ^{\infty}$-predictable,  $S$-integrable and satisfies \reff{noshortcont}. Define
\beqs
V_0^{\infty} &:=& \sup _{X\in \bar \Xc _c}\mathbb{E}[U(X_T)].
\enqs
Then  $ V_0^{\infty}=V_0^M$.
\end{Lemma}
{\bf Proof.} Since $\Xc ^S\subset \bar \Xc _c$, we obviously have $V^M_0\leq V^{\infty}_0$.
Now take some arbitrary 
$X$ $\in$ $\bar\Xc _c$ 
associated to a no-short sale trading strategy $H$, which is $\G ^{\infty}$-predictable.  Consider  the  $\G^S$-predictable projection of $H$: 
$\hat H_s$ $=$ $\E[H_s|\Gc_{s^-}^S]$, $t\leq s\leq T$.  We then have
\beqs
\hat X_t &:=& \E[X_t|\Gc_t^S] \; = \;  X_0 + \int_0^t \hat H_u dS_u, \;\;\; 0 \leq t \leq T. 
\enqs
This means that the process $\hat X$ lies in $\Xc^{S}$.  Since $U$ is concave, we get by the law of iterated conditional expectations and Jensen's inequality
\beqs
\E[U(X_T)] & = & \E \Big[ \E[U(X_T)| \Gc_T^S] \Big] 
 \leq  \E\Big[ U\Big( E[X_T| \Gc_T^S]  \Big) \Big] \;  = \; \E[U(\hat X_T)]  \; \leq \; V^M_0. 
\enqs
We conclude from the arbitrariness of $X$ in  $\bar\Xc _c$. 
\ep

\vspace{2mm}

\noindent {\bf Step 3}:  Once we prove the step above  and transform the Merton problem in a utility maximization problem with no short-sale constraints under the filtration   $\mathbb{G}^{\infty}$, we can basically follow the arguments in Theorems 3.1, 3.3 and 4.1 in  \cite{KardarasPlaten}, to finish the proof.

\vspace{1mm}

\noindent {\bf Proof of Theorem \ref{asymptotic}}.   
We  first represent  continuous time trading strategies with no short sale constraints  in terms of proportion of wealth, for the larger filtration $\G^{\infty}$.  For any continuous time wealth process $X$ $\in$ $\bar\Xc _c$ associated to a trading strategy $H$ satisfying \reff{noshortcont}, we still  denote  
 by  $\pi_t$ $=$ $H_t S_{t^-}/X_{t-}$,  and notice that the process  $(\pi _t)_{0\leq t\leq T}$ is valued in $[0,1]$ by 
\reff{noshortcont}.  We also denote by $X^{(\pi)}$ the process defined by \reff{pi}
 and  define  $\Dc^\infty$  to be the  set of all $\G ^{\infty}$-predictable processes $\pi$ 
 valued in $[0,1]$. It is then clear that 
 \beqs \label{vmerpi}
V_0^{\infty} &=& \sup_{\pi\in\Dc^\infty} \E[U(X_T^{(\pi)})].
 \enqs
 Using Lemma \ref{verification} and Lemma \ref{lemmer} we have, for each intensity function $\lambda$ that
 \beq \label{inegvM} 
 V_0^\lambda=V_0^{\lambda,c} \leq V_0^{\infty}=V_0^M.
 \enq
 Let now $\pi \in \Dc (S)$ $\subset$ $\Dc^\infty$  be a (proportional) trading strategy in \reff{defvMer} and let $(\lambda_k)_k$ a sequence of intensity functions as in Theorem \ref{asymptotic}.  We will follow the arguments in \cite{KardarasPlaten} to approximate this continuous time trading strategy by simple strategies  $\alpha^k\in \Ac ^{\lambda_k}_c$, which are discrete, but \emph{ use information from continuous  observations}.
  First, according to Lemma 3.4 and 3.5 in \cite{KardarasPlaten}, there exists  a sequence $\pi ^m \in \Dc (S)$ such that
  each $\pi ^m$ is LCRL (left continuous with right limits) and such that
  \beqs 
  uc\P -\lim _{m\rightarrow \infty} X^{(\pi ^m)}= X^{(\pi)},  
  \enqs
  so, in order to approximate $\pi$ we can actually assume it is LCRL. Here, by $uc\P$-convergence, we  mean the usual convergence of processes in probability, uniformly on compact time-sets. 
   
 Let us  check that our sequence of stopping times satisfies the condition $\sup_{n} \left|\tau_{n+1}^{k} - \tau_{n}^{k}\right| \rightarrow 0$ a.s., $k \rightarrow \infty$. Take a subdivision $0 = t_{0} < \ldots < t_{i} < \ldots < t_{M} = T$ of $[0,T]$ such that $\left| t_{i+1} - t_{i} \right| \leq \epsilon/2$, for all $i$. We then have :
\beqs
\P \Big[\sup_{n} \left|\tau_{n+1}^{k} - \tau_{n}^{k}\right| > \epsilon \Big] &\leq&
 \sum_{i=0}^{M-1} \P \Big[ \exists n, \tau_{n}^{k} \leq t_{i} < t_{i+1} \leq \tau_{n+1}^{k} \Big] \\
 &=& \sum_{i=0}^{M-1} \P \Big[ N_{t_{i}}^{\lambda_{k}}=N_{t_{i+1}}^{\lambda_{k}} \Big] \\
 &=& \sum_{i=0}^{M-1} \exp\Big( -\int _{t_{i}}^{t_{i+1}} \lambda _k (u)\,du\Big).
\enqs
By Borel-Cantelli and \reff{infintens}, we deduce that 
\beq \label{Borel}
\P \Big[ \limsup_{k} \sup_{n} \left|\tau_{n+1}^{k} - \tau_{n}^{k}\right| > \epsilon \Big] =& &0,\ \ \ \ (\forall)\ \varepsilon >0.
 \enq
 
 Next,   let us  define 
 \beqs \pi ^k _n= \pi _{\tau ^{\lambda ^k}_n+},  \ \ \ n\geq 0,
 \enqs
 where $\pi _{t+}=\lim_ {u\searrow t}\pi _t$.
 Because the filtration $\G ^{\lambda ^k}$ satisfies the usual conditions  we have that the process $(\pi _{t+})_{0\leq t\leq T}$ is optional with respect to $\G ^{\lambda_k}$ for each $k$. Since, in addition,  $\tau ^{\lambda_k}_n$ are stopping times with respect to 
 $\G ^{\lambda_k}$ we obtain that, for each $k$,
 \beq \label{adapted}
 \pi ^k _n \in \mathcal{G}^{\lambda ^k}_{\tau^{\lambda_k} _n} ,\ \ \ \ \ k\geq 0.
  \enq
  Therefore, if we define, for each fixed $k$ the discrete-time wealth process by 
  \beq
X_{\tau ^{\lambda_k}_{n+1}} &=& X_{\tau ^{\lambda_k}_{n}}(1 +   \pi ^k _nZ_{n+1}), \;\;\; n \geq 0,
\enq
and denote by   $\alpha ^k_n= \pi ^k _n X_{\tau ^{\lambda_k}_{n}}$
we have 
$  \alpha ^k=(\alpha ^k_n)_{n\geq 0}\in \Ac_c ^{\lambda_k}.
$  
To each of the above defined $\alpha^k$, we can associate by Remark \ref{remcontX} a continuous time simple integrand $H^k$ which is 
$\G ^{\lambda_k}$-predictable and the continuous time wealth  process $(X^k_t)_{0\leq t\leq T}$.
The fundamental observation is now that all $H^k$ are predictable with respect to the same ``large" filtration $\G ^{\infty}$. Using this universal filtration, we can now
follow the proof of Theorem 3.1 in \cite{KardarasPlaten},  which actually  works for stochastic partitions under condition \reff{Borel},  
to conclude that 
\beqs X^{(\pi)}=uc\P-\lim _{k\rightarrow \infty} X^k.
\enqs
Therefore, we can approximate any continuous time strategy in the Merton problem \reff{defvMer}  by simple trading strategies  
$\alpha ^k\in \Ac^{\lambda _k}_c$. The rather obvious details on how approximation of strategies leads to approximation of optimal expected utility are identical to the arguments in \cite{KardarasPlaten} Section 4, and are omitted: this means that for all $\pi$ $\in$ $\Dc(S)$, $\E[U(X_T^{(\pi)})]$ $=$ 
$\lim_{k} \E[U(X_T^k)]$, and so $V_0^M$ $\leq$ $\liminf_k V_0^{\lambda_k,c}$. Together with \reff{inegvM}, this concludes 
the proof of Theorem \ref{asymptotic}. 
\ep


\setcounter{equation}{0} \setcounter{Assumption}{0}
\setcounter{Theorem}{0} \setcounter{Proposition}{0}
\setcounter{Corollary}{0} \setcounter{Lemma}{0}
\setcounter{Definition}{0} \setcounter{Remark}{0}


\renewcommand{\theDefinition}{A.\arabic{Definition}}
\renewcommand{\theTheorem}{A.\arabic{Theorem}}
\renewcommand{\theLemma}{A.\arabic{Lemma}}
\renewcommand{\theRemark}{A.\arabic{Remark}}
\renewcommand{\theequation}{A.\arabic{equation}}
\setcounter{equation}{0}
\setcounter{Lemma}{0}
\setcounter{Remark}{0}
 
 \bibliographystyle{amsplain}

\begin{thebibliography}{99}

\bibitem{bershr78} Bertsekas D. and S. Shreve (1978): Stochastic optimal control: the discrete-time case, Academic Press. 


\bibitem{cregozphatan08} Cretarola A., Gozzi F., Pham H. and P. Tankov (2008): ``Optimal consumption policies in illiquid markets", 
to appear in {\it Finance and Stochastics}. 


\bibitem{jacshi03} Jacod J. and A. Shiryaev (2003): Limit theorems for stochastic processes, 2nd edition, Springer Verlag. 


\bibitem{KardarasPlaten} Kardaras, K. and E.  Platen (2008):  ``Multiplicative approximation of wealth processes involving no-short-sale strategies via simple trading", preprint, http://arxiv.org/abs/0812.0033


\bibitem{mat06} Matsumoto K. (2006): ``Optimal portfolio of low liquid assets with a log-utility function", 
{\it Finance and Stochastics}, {\bf 10}, 121-145. 

\bibitem{phatan08} Pham H. and P. Tankov (2008): ``A model of optimal consumption under liquidity risk with random
trading times'', {\it Mathematical Finance}, {\bf 18}, 613-627. 


\bibitem{rogzan02} Rogers C. and O. Zane (2002): ``A simple model of liquidity effects",  in 
{\it Advances in Finance and Stochastics: Essays in Honour of Dieter Sondermann},  
eds. K. Sandmann and P. Schoenbucher, pp 161--176. 


\end{thebibliography}

\end{document}